\documentclass[journal]{IEEEtran}

\usepackage{amsfonts}
\usepackage{amssymb}
\usepackage{stfloats}
\usepackage{cite}
\usepackage{graphicx}
\usepackage{epstopdf}
\usepackage{psfrag}
\usepackage{subfigure}
\usepackage{amsmath}
\usepackage{array}
\usepackage{stfloats}
\usepackage{multirow}
\usepackage{color}
\usepackage{booktabs}
\usepackage{url}
\usepackage{graphicx}
\usepackage{enumitem}
\usepackage{lipsum} 
\usepackage{float}
\usepackage{marvosym} 
\usepackage{url}
\usepackage{hyperref}
\usepackage{makecell}
\usepackage{bm}
\usepackage{eso-pic}
\usepackage[linesnumbered,ruled,vlined]{algorithm2e}
\usepackage{cases}

\begin{document}




\title{TREE: Token-Responsive Energy Efficiency Framework For Green AI-Integrated 6G Networks}

\author{Tao~Yu, Kaixuan~Huang, Tengsheng~Wang, Jihong~Li, Shunqing~Zhang,~\IEEEmembership{Senior Member, IEEE}, Shuangfeng~Han, Xiaoyun~Wang, Qunsong~Zeng,~\IEEEmembership{Member, IEEE}, Kaibin~Huang,~\IEEEmembership{Fellow, IEEE} and Vincent~K.~N.~Lau,~\IEEEmembership{Fellow, IEEE}

\thanks{Tao~Yu, Kaixuan~Huang, Tengsheng~Wang, Jihong~Li, Shunqing~Zhang (corresponding author) are with School of Communication and Information Engineering, Shanghai University, Shanghai, 200444, China (e-mails: \{yu\_tao, xuan1999, tengswang, tomlijiong, shunqing\}@shu.edu.cn).


Shuangfeng~Han and Xiaoyun~Wang are with China Mobile, Beijing, 100032, China (e-mail: \{hanshuangfeng, wangxiaoyun\}@chinamobile.com).

Qunsong Zeng and Kaibin Huang are with the Department of Electrical and Electronic Engineering, The University of Hong Kong, Hong Kong (e-mail: \{qszeng, huangkb\}@eee.hku.hk).

Vincent K. N. Lau is with the Department of ECE, The Hong Kong University of Science and Technology, Hong Kong. (e-mail: eeknlau@ust.hk.)
}
}

\markboth{Journal of \LaTeX\ Class Files,~Vol.~14, No.~8, August~2021}%
{Shell \MakeLowercase{\textit{et al.}}: A Sample Article Using IEEEtran.cls for IEEE Journals}


\maketitle

\begin{abstract}
As wireless networks evolve toward AI-integrated intelligence, conventional energy-efficiency metrics fail to capture the value of AI tasks. In this paper, we propose a novel EE metric called Token-Responsive Energy Efficiency (TREE), which incorporates the token throughput of large models as network utility carriers into the system utility. Based on this metric, we analyze the design principles of AI-integrated 6G networks from the perspective of three critical AI elements, namely computing power, model and data. Case studies validate TREE's unique capability to expose energy-service asymmetries in hybrid traffic scenarios where conventional metrics prove inadequate.  Although it is impossible to determine every design detail of AI-integrated 6G network at current time, we believe that the proposed TREE based framework will help the network operators to quantify the operating energy cost of AI services and continue to evolve towards sustainable 6G networks.
\end{abstract}


\section{Introduction}
\label{sect:intro}

The concept of AI-integrated Sixth-Generation (6G) networks has garnered significant attention in recent years, driven by the exponential growth of intelligent applications and the demand for seamless integration of artificial intelligence (AI) into wireless communication systems \cite{cui2025overview}. As outlined in the IMT-2030 white papers, the vision for 6G emphasizes ubiquitous intelligence, hyper-connectivity, and sustainability, with AI serving as a core enabler for autonomous network management, real-time decision-making, and dynamic resource allocation \cite{nagaraj2024demystifying}. Key driving forces include the need to support bandwidth-intensive applications (e.g., augmented reality, autonomous systems), massive Internet of Things (IoT) deployments, and new AI services such as dialogue systems and video generation. Emerging scenarios such as smart cities, industrial automation, and immersive Extended Reality (XR) further underscore the necessity of embedding AI natively into 6G architectures to achieve context-aware, self-optimizing networks for hybrid AI and data services.

Current research on AI-integrated 6G networks predominantly revolves around a cloud-edge-end architecture, designed to balance computational load, reduce latency, and enhance energy efficiency. For edge learning, techniques like federated learning, model splitting, and Low-Rank Adaptation (LoRA) based fine-tuning have been explored to enable distributed AI training while preserving data privacy and reducing communication overhead \cite{xu2023edge}. For edge inference, advancements include lightweight model deployment, real-time inference acceleration, and adaptive model compression to address resource constraints at edge devices \cite{yao2025energy}. Resource allocation and task offloading strategies, such as dynamic computation offloading and energy-aware scheduling, aim to harmonize AI workloads with network capabilities. Considering the distributed computing power at the network edge, some works focused on balancing computational load and reducing latency by decomposing AI workloads across heterogeneous network devices \cite{cai2024task}.

While significant progress has been made in architectural design and optimization techniques, the energy sustainability of AI-integrated 6G networks presents a critical challenge that demands urgent attention. First, from a energy consumption perspective, the development of energy-efficient AI-integrated 6G architectures requires systematic investigation of holistic energy consumption patterns and identification of key influencing factors across network components. Second, current energy efficiency studies for 6G networks predominantly adhere to conventional energy efficiency (EE) metric, typically defined as network throughput per unit energy consumption, which is not suitable for the AI services characterized by high computational energy consumption and token based user utility. There exists a pressing need to establish unified metrics that comprehensively quantify the system utility per joule of energy expenditure, particularly for new AI-enhanced services. These gaps hinder the development of sustainable 6G networks capable of supporting next-generation AI applications without exacerbating environmental impacts.

Critically, the economic landscape of network services is undergoing a paradigm shift. While traditional data services remain foundational, the explosive growth of generative AI services is fundamentally altering value creation dynamics. Industry data reveals a staggering valuation disparity: traditional data services yield approximately $4.26 \times 10^{-10}$ USD/bit in high-fee markets like the Netherlands \cite{fewer_operators}, whereas leading AI services such as ChatGPT-4 generate about $3.0 \times 10^{-5}$ USD/token \cite{akpan2024attention} – representing a nearly 70,000-fold revenue differential per unit. According to McKinsey, explosive growth in token processing demand could add $2.6\sim4.4$ trillion USD in value to the global economy each year \cite{chui2023economic}. These trends unequivocally demonstrate that tokens are rapidly becoming primary revenue drivers in next-generation networks.

This profound economic asymmetry between bits and tokens underscores a fundamental limitation of conventional bit-centric EE metrics: they fail to capture the vastly different economic value generated per unit of energy consumed for AI services. Consequently, there exists a pressing need to establish unified metrics that bridge AI service valuation with network energy expenditure, enabling operators to accurately quantify the economic efficiency (utility-per-joule) of AI-integrated networks. This economic imperative, combined with the technical limitations mentioned earlier, necessitates a novel framework.

In this paper, we introduce Token-Responsive Energy Efficiency (TREE), a novel metric that redefines energy efficiency to account for AI-specific utility in AI-integrated 6G networks. Departing from conventional EE metrics, TREE quantifies energy efficiency as the ratio of network utility (measured in AI-related tokens and transmission bits) to energy consumption, which directly capture the computational intensity and the value of AI tasks with token-level granularity. On top of that, we systematically derive design principles for AI-integrated 6G architectures through the lens of AI foundational triad including computing power, model and data. In addition, we  validate TREE's unique capability to expose energy-service asymmetries under hybrid traffic scenarios. The proposed metric and framework will help to eliminate the technical uncertainties and aggregate the efforts towards key breakthrough innovations for AI-integrated 6G networks.

\section{TREE Metric}

\begin{figure}[t]
\centering\includegraphics[height=10cm,width=8.5cm]{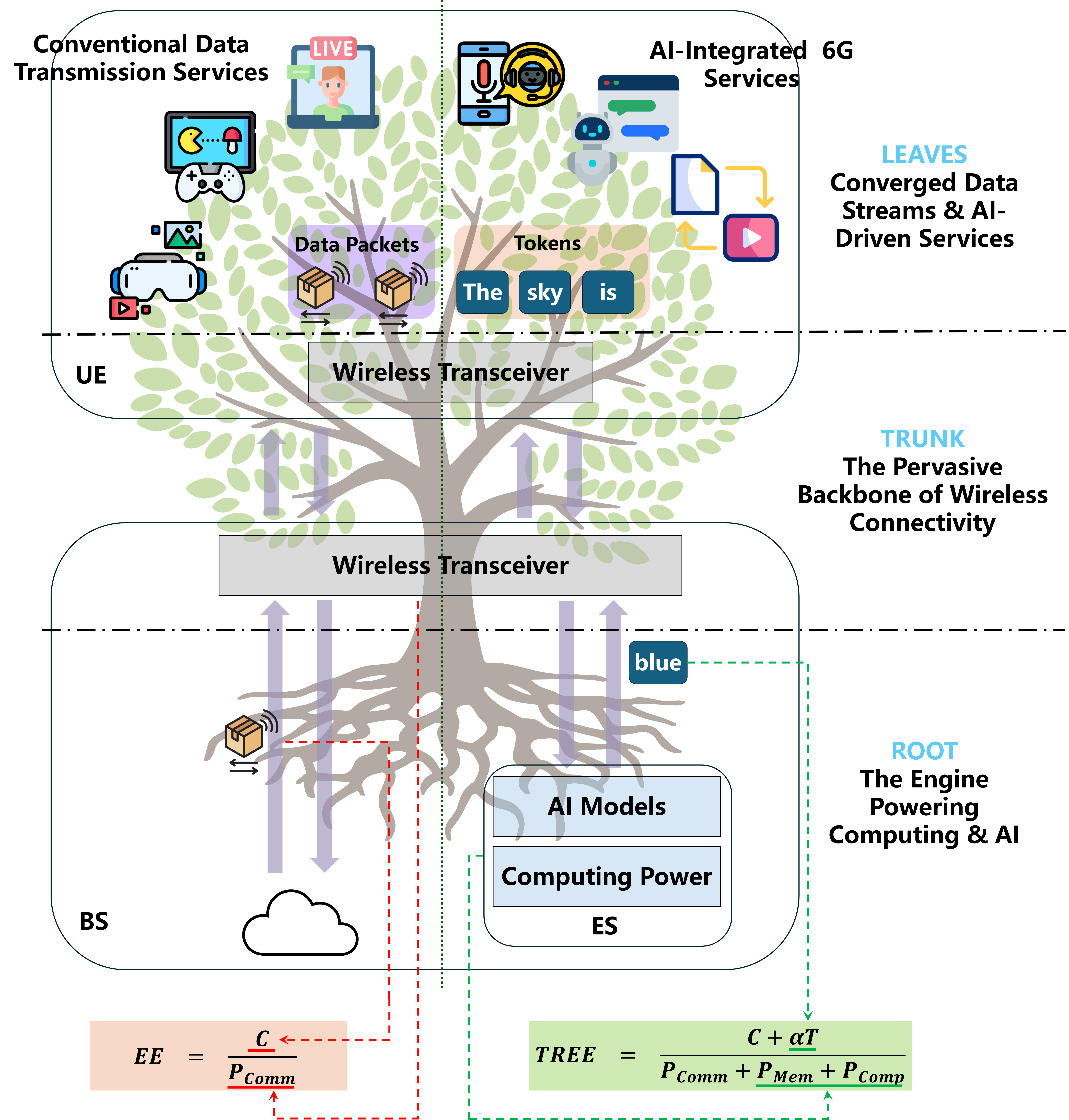}
\caption{ Comparative illustration of conventional EE vs. TREE framework. The left half depicts the traditional data packet flow, which aligns with EE's core definition: the ratio of bit utility to power consumption. In contrast, the right half illustrates the flow of AI services, which leverages edge or cloud computing resources and utilizes wireless communications as a channel to deliver novel token-based AI services. Consequently, while EE focuses solely on bit utility and power, TREE also incorporates token throughput, along with the power consumption associated with computation and memory. }
\label{fig:tree_def}
\end{figure}

The transition from conventional bit-centric wireless networks to AI-integrated 6G networks necessitates a fundamental redefinition of utility quantification. While conventional EE metrics focus on data throughput, they neglect the computational and economic value embedded in AI-generated outputs, i.e., LLM tokens. Tokens, intrinsically linked to computational workflows and directly indicative of service quality, establish themselves as a novel utility carrier. This paradigm shift elevates tokens to the cornerstone of TREE, providing a unified framework for evaluating and incentivizing energy-efficient designs in AI-integrated 6G networks.

\subsection{Why tokens?}

In the context of large language models (LLMs) and generative AI services central to AI-integrated 6G networks, tokens represent the fundamental atomic units of processing. They are discrete symbols derived from segmenting raw input data (text, image patches, audio features) using a predefined vocabulary or tokenizer. Each token corresponds to a meaningful subword, character, or contextual element within the model's framework. During both input ingestion and output generation phases, these tokens serve as the primary computational currency, sequentially processed by the model's neural architecture (e.g., transformer layers) to extract semantic meaning, perform reasoning, and ultimately produce intelligible AI-generated content. The rate and efficiency of token processing directly govern the perceived quality, latency, and utility of AI services delivered over the network.

The selection of tokens as the utility carrier for energy efficiency evaluation in AI-integrated 6G networks stems from their intrinsic alignment with the operational and economic dynamics of AI services. In traditional wireless networks, energy efficiency metrics predominantly focus on bits-per-joule ratios, reflecting the value of data transmission. However, this approach becomes inadequate for AI-integrated networks, where computational workflows generate value through token-based throughput rather than raw bit throughput. Tokens, as the atomic units of semantic meaning in LLM workflows, directly correlate with the quality, latency, and utility of AI services. For instance, each token generated in autoregressive decoding phases (e.g., ChatGPT-style responses) represents a quantifiable increment of service utility, whether for real-time translation, context-aware recommendations, or video generation. By adopting token throughput as utility, operators gain a granular metric that bridges AI task performance, user-perceived quality, and energy consumption.

Fundamentally, tokens serve as a natural intermediary between AI computational intensity and network resource allocation. Unlike conventional bit-based throughput, tokens inherently encapsulate the semantic density of AI outputs. For example, a single token in modern tokenizers may represent subwords, characters, or context-aware phrases, balancing computational efficiency with linguistic expressiveness. This granularity enables precise quantification of energy-per-service-unit costs. During inference phases, the iterative nature of token generation—where each token depends on preceding context—introduces unique energy consumption patterns. These patterns are shaped by factors such as prefilling latency, attention mechanism complexity, and dynamic batching strategies. By quantifying tokens rather than bits, the TREE metric directly addresses the computational asymmetry between AI tasks and legacy traffic, where energy consumption scales with tokens rather than bits.

Moreover, tokens provide a universal framework for cross-layer optimization in AI-integrated architectures. At the algorithmic level, token generation workflows directly influence network resource demands. Techniques like dynamic cache management and sparse attention mechanisms optimize token throughput while minimizing memory access energy. At the infrastructure level, token-aware resource allocation (e.g., adaptive spectrum scheduling or computation offloading) ensures that energy expenditure aligns with service value. This universality extends to heterogeneous AI workloads: whether deploying lightweight models at the edge for real-time anomaly detection or orchestrating massive LLMs in the cloud for context-aware network optimization, token-centric evaluation unifies energy efficiency analysis across diverse scenarios.

From an economic perspective, token-based utility measurement aligns with emerging token economies in AI service markets. As 6G operators transition toward value-driven resource pricing models, tokens offer a tangible unit for quantifying service operational costs and incentivizing energy-efficient behaviors. For instance, token throughput can inform dynamic pricing mechanisms where users or applications pay premiums for low-latency, high-token-throughput services. Simultaneously, operators can optimize energy-per-token costs through hardware-software co-design such as hybrid precision computing or pipeline-parallel processing. By embedding tokens into energy efficiency frameworks, 6G networks can foster sustainable growth, ensuring that the exponential demand for AI services does not outpace advancements in energy-aware infrastructure design.

\subsection{Definition of TREE}

Based on the above discussions, we summarize the definition of TREE as,
\begin{eqnarray}
\label{eqn:def_tree}
\eta_{TREE} &=& \frac{C + \alpha T}{P_{tot}}  
\end{eqnarray}
where $C$ denotes conventional data throughput in bits, $T$ represents the throughput of AI-related tokens, i.e. the number of input or output tokens per unit time. $P_{tot}$ is the total energy consumption, and $\alpha$ serves as a normalization factor converting tokens to bit-equivalent utility. As established in the Section~\ref{sect:intro}, the vastly different economic value generated per bit versus per token necessitates this factor to enable a unified energy efficiency metric that reflects the true economic utility delivered per joule of energy consumed. In fact, the TREE framework remains fully applicable to semantic communication paradigms where utility is quantified through semantic information \cite{luo2022semantic}, as it naturally accommodates both the value density of semantic content and the associated nonlinear power consumption.

As shown in Fig.~\ref{fig:tree_def}, the TREE metric introduces three critical advancements over conventional EE metric. First, it unifies the valuation of communication and AI computational services by jointly considering bit transmission and token generation as network utility carriers. This dual-component numerator addresses the limitation of traditional bit/joule metrics that disregard the energy-value asymmetry in AI tasks. Second, the inclusion of $\alpha$ enables service-aware energy efficiency optimization. Operators can set $\alpha$ to prioritize low-latency AI services (e.g., XR object recognition) requiring high token throughput or balance it with Enhanced Mobile Broadband (eMBB) traffic demands. Third, $P_{tot}$ accounts for the full energy footprint across cloud-edge-end tiers, including computation, Input/Output (I/O), and transmission costs, which is necessary for holistic sustainability analysis in distributed AI workflows.

The proposed TREE offers unique benefits for 6G network design and operation. Technically, it provides a universal optimization target for heterogeneous deployments, for example, edge nodes can prioritize $T$ for localized AI services while macro-cells emphasize $C$ for broadband traffic, all within a unified energy efficiency framework. From an economic perspective, TREE quantifies the operational energy cost of AI services per generated token, enabling operators to develop token-driven resource pricing models.

\section{TREE oriented Design} \label{sect:TREE_oriented_Design}

A key implementation consideration for TREE lies in the measurement of $T$ and its energy attribution. Unlike bit transmission, token processing highly spans to AI foundational triad including computing power, model and data. In this section, we built a systematic framework from the above three perspectives and sorted out the design of green AI-integrated 6G network based on TREE.

\subsection{Computing Power Perspective}

The computing power perspective necessitates a new perspective in resource orchestration to align heterogeneous communication, computational and memory capabilities with TREE maximization. Considering a edge server (ES) with $N_{core}$ activating computing cores, the total power consumption is given as \cite{8716300},
\begin{eqnarray}
\label{eqn:total_power}
P_{tot} &=& \underbrace{ v_c^2 f_c \left (\sum_{k =1}^{N_{core}} \lambda^c_1 u_c^{(k)} + \lambda^c_2 \right) + v_c \lambda^c_0 }_{\text{ Computation power consumption, } P_{Comp}}  \nonumber \\
&+& \underbrace{v_m^2 f_m \left (\lambda^m_1 u_m + \lambda^m_2 \right) + v_m \lambda^m_0}_{\text{Memory power consumption, } P_{Mem}}  \nonumber \\
&+& \underbrace{\lambda^r  \sum_{j=1}^{N_{sub}} p_{t}^{(j)} + p_{c}}_{\text{Communication power consumption, } P_{Comm}}
\end{eqnarray}
where $v_c$ and $f_c$ are the voltage and the frequency of the Graphics Processing Unit (GPU)/Central Processing Unit (CPU), $u_c^{(k)}$ is the average utilization rate of the computation resource for $k$-th core, which is also called Model FLOPS Utilization (MFU) of AI models. $v_m$, $f_m$ and $u_m$ are the voltage, the frequency and the average utilization rate of the memory. $p_{t}^{(j)}$ is the wireless transmit power for subcarrier $j$ and $p_{c}$ is the total circuit power. $\lambda^c_0, \lambda^c_1, \lambda^c_2, \lambda^m_0, \lambda^m_1, \lambda^m_2 $ are the hardware specific parameters, associated to the characteristics of the underlying architecture, such as component total capacitance and leakage resistance. $\lambda^r$ signifies the amplification coefficient linked to the power amplifier efficiency.

Static power consumption, governed by hardware manufacturing processes and leakage currents, underscores the importance of joint communication-computation hardware co-design. Advanced chip fabrication techniques, such as monolithic three-dimensional integration and near-memory computing architectures, minimize leakage-related terms (i.e., $v_c \lambda_0^c$ and $v_m \lambda_0^m$) by reducing interconnect distances and optimizing transistor threshold voltages. Emerging paradigms like in-memory computing further mitigate static overhead by co-locating AI operations within memory arrays, eliminating energy-intensive data movement between processing and memory units. For wireless transceivers, innovations in reconfigurable Radio Frequency (RF) frontends and hybrid beamforming architectures harmonize static circuit power $p_{c}$ with dynamic amplification efficiency $\lambda^r$, enabling energy-proportional operation across diverse AI workload scenarios.

As for dynamic energy consumption, there are mainly two types of energy efficient schemes given as follows,
\begin{itemize}
    \item {\em Adaptive sleeping mechanism for functional components}. Through a monitoring \& sleep manner, the discontinuous reception (DRX) and transmission (DTX) techniques in the existing 5G network are already able to support long-cycle (such as 1000ms) and short-cycle (such as 20ms) circuit sleep, where the circuit wake-up time is proportional to the sleep depth. In order to accommodate the characteristics of different AI tasks, rapid state transitions is applied with frequent lightweight inferences, while sustained sleeping is applied with intermittent heavy computations. However, the interdependent nature of communication and computation in AI-integrated 6G pipelines introduces unique coordination challenges. 
    The frequency switching latency of computing units (5-500 ms \cite{velicka2025methodologygpufrequencyswitching}) could nullify energy savings if not synchronized with memory subsystem reactivation and communication resource preparation. Based on that, the design of adaptive sleeping mechanisms for AI-integrated 6G networks necessitates a hierarchical time-scale framework that harmonizes the heterogeneous wake-up latencies and energy-saving potentials across computational, memory, and communication subsystems. In this  hierarchical framework, a predictive pre-wakeup coordination that leverages token and bit workload characteristics is necessary to stagger the reactivation of compute cores, memory banks, and wireless subcarriers. For instance, when processing video generation tasks with predictable prefilling phases, edge servers can initiate memory subsystem wakeup during video frame reception, followed by gradual computational core activation aligned with decoder phase demands.
    
    \item {\em Dynamic management for multi-dimensional resources}. The cornerstone of energy-efficient resource management in AI-integrated 6G networks lies in navigating the nonlinear interdependencies between power consumption and utility under stringent resource constraints. Conventional wireless communication systems balance the Shannon capacity and transmit power through water-filling algorithms, but AI introduces additional layers of complexity. As delineated in \eqref{eqn:total_power}, computational and memory subsystems exhibit nonlinear power-frequency-voltage relationships, where GPU/CPU frequency scaling and voltage adjustments dynamically influence core utilization rates (e.g., MFU) and token generation throughput. This inherent nonlinearity necessitates joint optimization of multi-dimensional resources (voltage, frequency, bandwidth, transmit power) to maximize TREE. Recent advances in MFU modeling and AI task characterization \cite{pope2023efficiently} provide critical insights into workload-dependent energy profiles. By integrating Dynamic voltage and frequency scaling (DVFS) driven compute optimization with wireless channel-aware resource allocation, operators can dynamically adjust GPU/CPU/memory operating points and wireless transmit power to align token generation rates, data transmission rates with time-varying wireless conditions. 
\end{itemize}

These energy-efficient schemes operate synergistically to maximize the TREE metric. Specifically, the adaptive sleeping mechanism minimizes idle power consumption within the total energy expenditure $P_{tot}$ by dynamically deactivating underutilized hardware components during periods of low demand. Concurrently, dynamic resource management optimizes the conventional data throughput $C$ and AI token throughput $T$ by aligning computational resources (voltage, frequency, etc.) with the time-varying workload demands of both communication and AI services. These combined optimizations directly enhance $\eta_{TREE}$ by simultaneously reducing the denominator $P_{tot}$ and increasing the numerator $C+\alpha T$, thereby maximizing the system's utility-per-joule efficiency as defined by the TREE framework.

\subsection{Model Perspective}

\begin{figure}[t]
\centering\includegraphics[height=11.5cm,width=8.5cm]{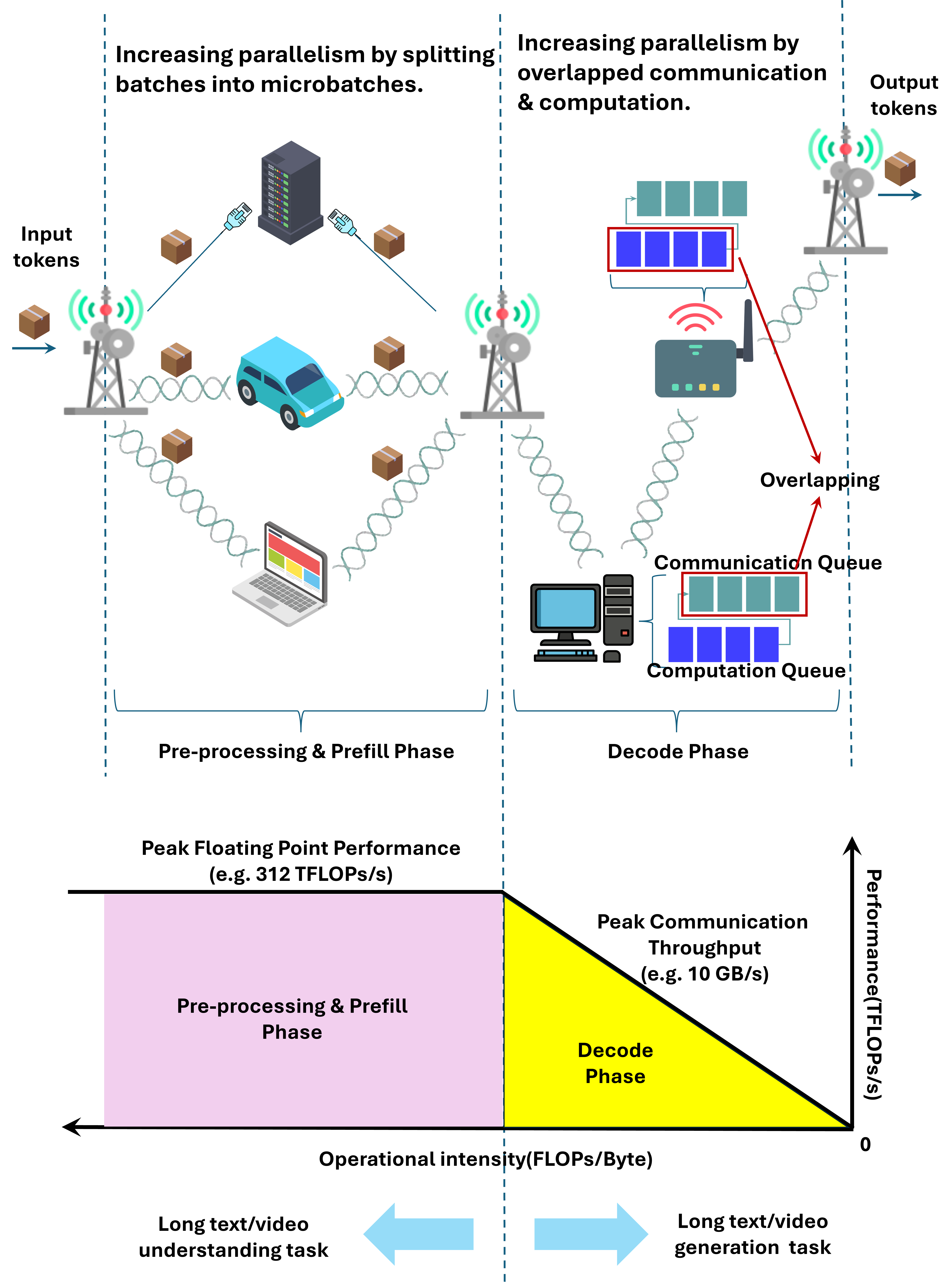}
\caption{ Token generation workflow in AI-integrated 6G networks. The links among these edge devices could be wireless or wired links. During the prefill phase, token throughput is enhanced by parallel processing of multiple microbatches across devices. In the decode phase, throughput is improved by overlapping computation and communication, reducing bubble time—idle periods in device computation due to communication or synchronization overhead. As indicated by the roofline model, performance is limited by computing power in the prefill phase and communication throughput in the decode phase.
} \label{fig:model_persp}
\end{figure}

The model architecture and inference workflow fundamentally govern the token generation efficiency and energy characteristics in AI-integrated 6G networks. Modern large AI models typically consist of three core modules including pre-processing module, encoder, and decoder, each corresponding to distinct computational phases with unique resource utilization patterns. In the pre-processing phase, tokenization and embedding operations dominate, characterized by sequential branch-intensive operations that heavily utilize CPU resources while leaving GPU accelerators underutilized. This computational asymmetry stems from the inherent nature of tokenizers that rely on conditional logic for subword segmentation, creating fundamental bottlenecks for hardware acceleration. The subsequent prefill phase transforms these token embeddings through massive parallel computations in encoder layers, leveraging GPU parallelism for simultaneous attention mechanism calculations across all input tokens. This phase exhibits computational scalability but demands significant energy expenditure proportional to the token sequence length and attention complexity. While mathematically equivalent to prefill computations, the decode phase's memory-bound nature emerges from its dependency on progressively growing key-value (KV) caches. Each generated token triggers read-modify-write operations on these caches, creating an energy profile dominated by memory subsystem activity rather than pure compute operations. This phase reveals a critical tradeoff between computational efficiency and memory access energy, particularly problematic for long-context interactions. Pipeline parallelism emerges as a critical design paradigm in distributed AI systems \cite{li2021terapipe}. As shown in Fig.~\ref{fig:model_persp}, using pipeline parallelism, the inherent computational intensity can be mitigated through microbatch partitioning across edge devices for pre-processing and prefill phases. While asynchronous pipeline execution across edge devices such as interleaved scheduling requires tight coordination to reduce the bubble time and maximize the token throughput.

The convergence of AI workflows with 6G network necessitates fundamental rethinking of wireless system design to align with the above token pipelines.

\begin{itemize}
    \item {\em Phase-aware protocol stack upgrade}. 6G protocol stack upgrade emerges as critical to harmonize computational phase characteristics with network resource management. Modern protocol stacks should embed AI processing phase identifiers (pre-processing/prefill/decode) in control plane messages, coupled with real-time CPU/GPU/memory utilization metrics. This enables joint Quality of Service (QoS) scheduling that dynamically prioritizes data or token based throughput, latency, and energy constraints during different phases. For pre-processing and prefill phases requiring microbatch partitioning and distributed matrix operations, the protocol stack must support tensor-parallel transmission through enhanced PDCP layer segmentation and MAC concatenation mechanisms. Conversely, decode phase demand tight temporal coordination across distributed pipeline stages where 6G native time-sensitive networking (TSN) capabilities synchronize token generation intervals with URLLC-grade minislot structures, minimizing pipeline bubbles through deterministic latency guarantees. 

    \item {\em Token/feature-oriented transmission mechanisms}. The 6G air interface also requires structural modifications to accommodate token-based processing pipelines. Since most of the transmitted data are the features of microbatches, pre-processing and prefill phases could employ over-the-air computation (AirComp) principles to reduce the total latency by processing data during transmission. With advanced signal proccessing technology, AirComp inherently perform neural network operations like AllReduce and AllGather through wireless signal superposition. In the decoding phase, the focus is on ensuring the transmission of high-value tokens. Therefore, semantic aware transmission schemes are encourged. For example, modulation and coding schemes (MCS) dynamically shall adapt to token entropy profiles. High-entropy tokens (e.g., novel information in dialog turns) employ QPSK with low-rate Low-Density Parity-Check (LDPC) codes for ultra-reliable delivery, while low-entropy tokens (e.g., conversational fillers) utilize 1024 Quadrature Amplitude Modulation (QAM) and high-rate polar codes to maximize token throughput. 
    
    \item {\em Task-centric resource pooling and slicing}. In order to cooperate with the above-mentioned token processing workflow, it is necessary to dynamically pool resources and abstract wireless spectrum, power resources, computing, and memory resources into programmable resource units. On top of that, a task-centric slicing framework can be constructed based on the famous roofline model \cite{ofenbeck2014applying}. For upstream-dominant tasks like long video understanding, network slices prioritize prefill-optimized configurations: wideband uplink (UL) channels paired with high-MFU GPU partitions and proactive KV cache preloading. Downstream-centric slices (e.g., real-time video generation) allocate decode-focused resources—low-latency Time Division Duplex (TDD) frames synchronized with pipeline stages, in-memory compute substrates for token finalization. A blockchain-backed resource ledger tracks token-generation energy costs across slices, enabling TREE-optimal workload placement through advanced algorithms that balance the asymmetry in computing power and  wireless channel.
\end{itemize} 

These design principles are intrinsically linked to the TREE metric through their direct impact on its constituent elements. Phase-aware protocol upgrades, such as embedding AI processing phase identifiers and enhancing coordination, significantly enhance token throughput $T$ by minimizing pipeline bubbles inherent in distributed token generation workflows, as depicted in Fig.~\ref{fig:model_persp}. Concurrently, token/feature-oriented transmission (e.g., leveraging AirComp) and task-centric resource pooling/slicing mechanisms boost the overall system utility $C+\alpha T$ by enabling more effective allocation of communication and computational resources. Crucially, these combined design efforts work synergistically to minimize the energy expenditure per token $P_{tot}/T$, thereby directly elevating the core TREE metric $\eta_{TREE}$ defined in \eqref{eqn:def_tree}.

\subsection{Data Flow Perspective}

The transition from single-user token processing workflows to multiple user equipments (UEs) and base stations (BSs) environment introduces unprecedented complexity in data flow orchestration. Multi-UEs/BSs interactions demand simultaneous deal with wireless channels, computational resource, and dynamic network topologies while preserving TREE objectives. At its core, this challenge stems from the interplay between two dimensions of heterogeneity, the wireless channel diversity across users/BSs and the computational request asymmetry in AI workloads. To harmonize these factors, we have to rethink and upgrade the current schemes.

\begin{itemize} 
\item {\em Dynamic batching combined multi-user scheduling}. Conventional wireless multi-user scheduling schemes such as proportional fairness or channel-state-information (CSI)-based prioritization merely rely on the wireless performance and ignore the token-based workflow. Conversely, dynamic batching is an advanced AI technique that dynamically processing multiple input sequences (may from multiple users) by adjusting batch sizes to maximize computational efficiency without wireless environment considerations. Therefore, the multi-user scheduling mechanisms in AI-integrated 6G network must evolve to integrate these two techniques to achieve TREE maximized multi-user token processing.  Specifically, when scheduling a group of users, the batch size (i.e., the number of scheduled users) should be determined to determine the most appropriate parallelism for the current resources, and then the optimal resource allocation plan for each batch should be determined. In the above process, considering the random fluctuations of the wireless channel, the transmission delay should be fully considered to reduce the bubble time of the entire workflow.

\item {\em AI workflow optimized multi-BS connectivity}. The integration of multi-BS connectivity mechanisms, particularly through Packet Data Convergence Protocol (PDCP) layer split transmission standardized in 5G/6G, offers potential for optimizing AI workflow execution in 6G networks. By enabling simultaneous data transmission through multiple BSs, PDCP layer split dynamically partitions and routes tokenized AI workloads through heterogeneous radio links, balancing computational load and maximizing the throughput. For the pre-process and prefill phase, microbatch tensor partitions are mapped to parallel radio bearers, enabling channel-aware resource allocation where high-reliability links handle critical tokens (e.g., decoder attention states) while high-capacity paths transport bulk embeddings. For decode phase with stringent synchronization requirements, this approach mitigates single-BS bottlenecks while exploiting spatial diversity to stabilize token throughput under mobility or channel fading. To fully harness this synergy, AI-integrated 6G protocols must embed dynamic coordination between PDCP flow control and computational state management, ensuring token generation continuity for TREE optimization.

\item {\em Task specialized mobility management}. The inherent mobility of UEs and movable BSs in 6G networks further complicates data flow management. These mobility induce rapid changes in network topology, dynamically altering the availability and proximity of communication and computing resources. Conventional wireless handover mechanisms, designed for data connectivity preservation, must be augmented with computational state migration protocols to maintain token-generation continuity. For example, when a user hand over between BSs, the serving edge server must transfer partial KV caches, tokenizer contexts, and microbatch states to the target BS’s edge node with millisecond synchronization. This demands tight integration of radio resource control (RRC) signaling with distributed computing frameworks, where handover triggers initiate parallelized checkpointing and prefetching of computational states. Furthermore, topology-aware resource pooling protocols dynamically reconfigure virtualized resource slices (spectrum, compute, memory) across moving BS clusters, ensuring efficient TREE performance despite geographical redistribution of users and infrastructure.  
\end{itemize} 

Dynamic batching combined with multi-user scheduling optimizes token throughput $T$ and conventional data throughput $C$ by adaptively grouping tasks to maximize GPU utilization while minimizing communication latency. AI workflow optimized multi-BS connectivity further reduces the total energy consumption $P_{tot}$ through load-balanced resource pooling. Task-specialized mobility management minimizes state migration overhead during handovers. These principles collectively increase the utility-per-joule ratio $\eta_{\text{TREE}} = (C + \alpha T) / P_{tot}$ under hybrid workloads by directly elevating the numerator $C + \alpha T$ while suppressing the denominator $P_{tot}$.

\begin{figure*}[t]
\centering\includegraphics[height=6cm,width=18cm]{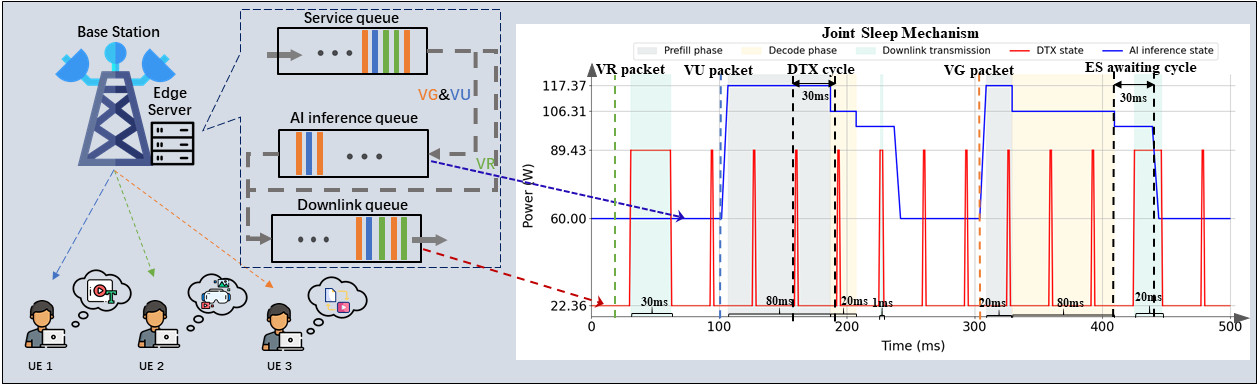}
\caption{ Downlink transmission process of Hybrid services, where the BS and ES employ a joint sleep mechanism to save energy. Specifically, the DTX technology employed by the BSs reduces static communication power consumption by discontinuously transmitting during idle intervals, while DVFS adopted by ESs dynamically adjusts the voltage and frequency based on real-time workload demands. The energy consumption difference between the prefill phase and the decode phase can be observed.}
\label{fig:joint_sleep}
\end{figure*}

\section{Case Study: Performance of Hybrid Services With Joint Sleep Mechanism}


To validate the proposed TREE metric's effectiveness in capturing the energy-service interplay within AI-integrated 6G networks, this case study examines a hybrid service scenario featuring three distinct service types representative of core AI-driven demands: Virtual Reality (VR), Video Understanding (VU), and Video Generation (VG). These services exhibit divergent resource dependencies. VR is communication-intensive and bypasses ES inference for high-bitrate downlink; VU is computation-intensive with dominant prefill-phase overhead; VG is memory-intensive due to significant decode-phase I/O operations. Building on the design principles in Section~\ref{sect:TREE_oriented_Design}, we implement a joint sleep mechanism where the BS employs DTX and the ES utilizes DVFS for energy conservation during idle periods, as illustrated in Fig.~\ref{fig:joint_sleep}. Specifically, DTX reduces static communication power consumption $p_{c}$ by discontinuously transmitting during idle intervals, while DVFS dynamically adjusts ES voltage $v_c$ and frequency $f_c$ based on real-time workload demands, optimizing the computational energy consumption $P_{comp}$. This synergy minimizes total energy consumption $P_{tot}$ while maintaining utility $C + \alpha T$, thereby maximizing $\eta_{TREE}$. For packet processing, VR packets proceed directly to the downlink queue, whereas VU and VG packets undergo ES inference. The ES upclocks upon request arrival to execute prefill and decode phases sequentially, then awaits potential downclocking if no new requests arrive. This configuration enables rigorous TREE performance assessment under varying packet arrival rates while highlighting resource contention and energy adaptation dynamics in hybrid AI-communication workloads. The detailed simulation configurations is listed in Tab.~\ref{tab:simu_para}.

\begin{table} [htpb] 
\centering 
\caption{Configurations}  
\label{tab:simu_para}
\footnotesize
\begin{tabular}{c | c }  

\toprule 
Parameters & Values \\

\midrule
BS & Micro BS with $1024$ subcarriers\\

\midrule
GPU & GeForce RTX 4090 \\

\midrule
\makecell{Communication power \\ consumption, $P_{Comm}$ } & \makecell{Transmit power, $1000$ mW  \\ RF chain power, $1000$ mW  \\
Fixed circuit power, $10000$ mW} \\

\midrule
\makecell{Memory power \\ consumption, $P_{Mem}$ } &\makecell{Voltage, $1.35$ V  \\ Frequency,  $10.50$ GHz }  \\

\midrule
\makecell{Computation power \\ consumption, $P_{Comp}$ } & \makecell{Voltage, $1$ V  \\ Frequency,  $2.52$ GHz } \\

\midrule 
DTX settings & \makecell{DTX cycle, $30$ ms  \\ Switching delay,  $1$ ms } \\

\midrule 
DVFS settings & \makecell{ES awaiting cycle, $30$ ms  \\ Switching delay,  $5$ ms \\ Sleep power , $60.00$ W \\ Inference power , $117.37$ W} \\

\midrule 
Traffic settings &\makecell{VR payload size,1.95 MByte/frame \\VG payload size, 1.5 MByte/frame \\VU payload size,40 kByte \\ }\\

\bottomrule
\end{tabular}  
\end{table}

\begin{figure*}[t]
\centering\includegraphics[height=12cm,width=18cm]{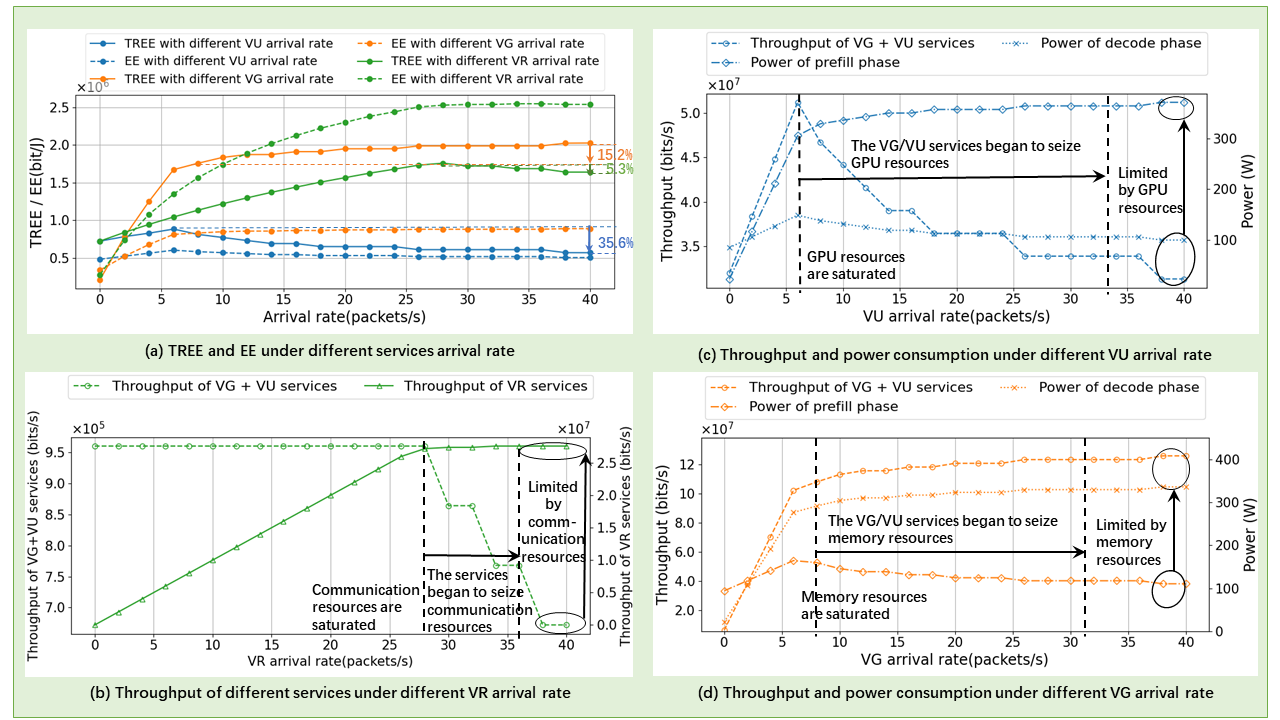}
\caption{ Performance evaluation of hybrid services under joint sleep mechanism. Unlike EE metric, TREE captures variations in utility and energy consumption resulting from the contention for communication, computing, and memory resources among three types of services, thereby more accurately reflecting the system-wide energy efficiency performance of AI-integrated 6G networks.} 
\label{fig:case_study}
\end{figure*}

In Fig.~\ref{fig:case_study}, we demonstrate network performance trends under varying packet arrival rates across distinct services with $\alpha = 10$. The principal findings can be summarized as follow.

\begin{itemize} 
\item {\em Performance under different VR packet rate}. As shown in Fig.~\ref{fig:case_study}(a) and Fig.~\ref{fig:case_study}(b), with VR packet arrival rate increases from low to saturation, both EE and TREE initially rise due to improved resource utilization. Beyond a critical threshold, VR traffic dominates communication resources, suppressing AI service throughput. While EE plateaus (reflecting saturated bit/joule efficiency), TREE exhibits sustained decline until stabilization. This divergence stems from TREE’s dual sensitivity: (1) reduced AI token throughput from resource competition, and (2) dynamic power reduction in idle compute/memory subsystems during AI workload suppression – a dimension EE inherently overlooks. The metric thus uniquely captures the energy-service asymmetry when communication-intensive services preempt AI workflows.

\item {\em Performance under different VU packet rate}. As shown in Fig.~\ref{fig:case_study}(a) and Fig.~\ref{fig:case_study}(c), increasing VU packet arrival rate initially boosts both EE and TREE through enhanced joint utilization of communication and computational resources. Upon reaching computational saturation, further rate escalation triggers contention between VU and VG services for limited GPU/memory resources. This contention suppresses total AI throughput due to the relatively small downlink data volume of VU services (e.g., object recognition results of VU service), driving synchronized declines in both metrics. Crucially, TREE exhibits amplified sensitivity due to its two-aspect response: (1) reduced service throughput, (2) idle-memory power savings from VU preemption – a nuanced energy-profile shift EE inherently overlooks. Stabilization occurs when VU traffic saturates service queues, validating TREE’s granular capture of cross-service computational interference.

\item {\em Performance under different VG packet rate}. As shown in Fig.~\ref{fig:case_study}(a) and Fig.~\ref{fig:case_study}(d), with VG packet arrival rate increases, both TREE and EE demonstrate monotonic growth until stabilizing at saturation, with TREE exhibiting significantly larger relative gains. This amplification originates from VG's inherent high-volume downlink data payloads (e.g., HD video frames) which substantially elevate utility-per-joule efficiency. Upon reaching memory saturation thresholds, VG contention suppresses VU service throughput, yet net AI throughput continues rising due to VG's data dominance. Crucially, TREE uniquely registers concurrent power reduction in prefill phases as VG preemption idles GPU resources – an energy dynamic invisible to conventional EE metrics. Both metrics stabilize when VG traffic saturates service queues, validating TREE's granular responsiveness to memory-bound service interactions.

\end{itemize}

The TREE metric uniquely captures the critical interplay between performance asymmetry and power dynamics inherent in hybrid service environments. It simultaneously quantifies the service performance disparity—where conventional throughput-centric services contend with AI-driven workloads (e.g., VR traffic preempting communication resources)—and the consequent total system power variations induced by cross-service competition for computation or memory resources (such as the fluctuating power profiles observed during VU and VG mutual preemption of prefill and decode phases). This dual quantification establishes TREE as a robust metric essential for guiding the sustainable design of future 6G networks.

\section{Potential Challenges}

While the TREE framework and its architectural principles establish a solid foundation for energy-efficient AI-integrated 6G networks, practical deployment faces systemic challenges in the following dimensions.

\begin{itemize} 
\item {\em Distributed power consumption ledgers and audit}. A significant challenge in AI-integrated 6G networks due to the need to reconcile heterogeneous energy accounting mechanisms across cloud-edge-end infrastructures. A critical barrier is the absence of standardized methodologies for attributing energy consumption to shared computational resources, particularly in distributed AI workflows, which introduces fundamental measurement inconsistencies across systems. While current research demonstrates progress towards resilient edge-cloud frameworks integrating lightweight auditing with blockchain-anchored checksums for data integrity, and distributed ledger technology (DLT)-integrated fault management systems, practical deployment faces hurdles. The temporal decoupling between token-generation workflows and their associated energy expenditure complicates real-time auditing. To address this, novel temporal alignment protocols must be developed to synchronize computational state transitions with power measurement intervals, ensuring accurate energy-per-token attribution within the stringent latency constraints of 6G operations.


\item {\em Hardware co-design for AI and communications}. Traditional AI accelerators (e.g., GPUs) prioritize floating-point operations per watt but lack native support for real-time spectrum awareness and adaptive beamforming, while baseband processors optimized for massive Multiple-Input Multiple-Output (MIMO) spectral efficiency struggle with the memory-bound nature of transformer-based inference, creating a significant architectural dichotomy that results in energy inefficiencies at the hardware-software interface; however, current research \cite{tao2022algorithm}  demonstrates the effectiveness of algorithm-architecture co-design in bridging this gap, achieving substantial gains in efficiency and throughput for both key channel decoders (e.g., polar/LDPC codes) and neural networks through architectural innovations like task-specific parallelism and sparsity exploitation, with interdisciplinary advances further showcasing the potential of unified designs by using AI modules to enhance traditional communication algorithms; nevertheless, core challenges remain including enabling microsecond-scale hardware reconfigurability to align AI token generation dynamics with communication scheduling cycles, holistically managing heterogeneous computational demands across domains, and achieving scalable energy efficiency for increasingly complex combined workloads beyond today's specialized accelerators.

\item {\em Standardization}. Standardizing TREE metrics and AI-integrated architectures faces significant coordination hurdles due to fragmented stakeholder interests and legacy frameworks. Current standardization bodies, e.g., 3rd Generation Partnership Project (3GPP), International Telecommunication Union (ITU), European Telecommunications Standards Institute (ETSI), focus predominantly on communication-centric energy efficiency metrics, with limited mechanisms to harmonize token-based utility valuation across heterogeneous AI services. Compounding this challenge, regional regulatory policies and divergent industry priorities further impede consensus on unified evaluation methodologies.

\end{itemize}

\section{Conclusion}

The TREE framework bridges the critical gap in energy efficiency evaluation for AI-integrated 6G networks by unifying token-based AI utility with conventional bit-oriented metrics. By formulating foundational design principles through systematic analysis of three critical AI pillars—computing power, model architecture, and data flow—this work establishes a methodology to harmonize token-generation workflows with energy-aware resource orchestration. Case studies on hybrid services validate TREE’s unique capability to expose energy-service asymmetries in scenarios where conventional metrics fail, demonstrating its effectiveness in evaluating computational intensity, memory access patterns, and wireless capacity. However, practical deployment faces challenges in distributed power auditing, hardware co-design for joint communication-computation, and standardization of token-centric valuation. As 6G evolves toward AI-native architectures, TREE provides an essential paradigm for quantifying and optimizing the sustainability of intelligent services, guiding operators toward reconciling exponential AI workload growth with global carbon neutrality objectives.

\section*{Acknowledgments}

This work was partly supported by the National Key Research and Development Program of China under Grant 2022YFB2902304, and the Science and Technology Commission Foundation of Shanghai under Grants 24DP1500703. The work by Q. Zeng was supported by the Natural Science Foundation of Guangdong Province under Grant 2025A1515011747.

\bibliographystyle{IEEEtran}
\bibliography{IEEEfull,references}

\end{document}